# Phononic frequency comb via intrinsic three-wave mixing

Authors: Adarsh Ganesan[1], Cuong Do[1], Ashwin Seshia[1]

1. Nanoscience Centre, University of Cambridge, Cambridge, UK

**Optical frequency combs [1-8] have resulted in significant advances in optical frequency metrology and found wide application to precise physical measurements [1-4, 9] and molecular fingerprinting [8]. A direct analogue of frequency combs in the phononic or acoustic domain has not been reported to date. In this letter, we report the first clear experimental evidence for a phononic frequency comb. In contrast to the Kerr nonlinearity [10] in optical frequency comb formation, the phononic frequency comb is generated through the intrinsic coupling of a driven phonon mode with an auto-parametrically excited sub-harmonic mode [16]. Through systematic experiments at different drive frequencies and amplitudes, we portray the well-connected process of phononic frequency comb formation and define attributes to control the features [17-18] associated with comb formation in such a system. Further, the interplay between these nonlinear resonances and the well-known Duffing phenomenon [12-14] is also observed. The presented pathway for phononic frequency comb formation finds general relevance to other nonlinear systems in both classical and quantum domains.**

A frequency comb consists of a series of equally spaced discrete frequencies. In recent years, optical frequency combs [1-8] have emerged as a potential toolset spanning diverse applications ranging from frequency metrology [1-4] to molecular fingerprinting [8]. Specifically the ability to precisely define the frequency spacing between frequency markers and align these measurements with microwave sources through the comb generation process has led to a number of physical measurements [4] requiring very high accuracy including the observation of gravitational waves [9]. Optical frequency combs have been generated by using the comb-like mode structure of mode-locked lasers and more recently through the interaction of continuous-wave lasers with high Q toroidal optical microresonators mediated via the Kerr non-linearity [10].

Despite the analogies between phonons and photons, a direct analogue for an optical comb in the phononic domain has not been observed and comb generation process is thought to be largely limited by the nonlinear dispersion relations of phonons. However, theoretical work [11] has recently demonstrated the possibility for generation of frequency combs in a phononic system represented by Fermi-Pasta-Ulam α (FPU-α) chains [15] where the dispersion relationship does not play a role. In this letter, we report the first experimental confirmation of a phononic frequency comb in a microfabricated structure bearing similar traits predicted by numerical simulations performed on a FPU-α chain [11]. Additionally, our resonator also captures the onset of Duffing nonlinear mechanism [12-14] and its interference with the nominal phononic comb.

The non-linear three wave mixing mechanism resulting in the generation of frequency combs is theoretically facilitated through the excitation of non-linear resonances of various orders. Specifically, in Direct Nonlinear Resonance (DNR) as termed in [11], the interaction between the eigenmode and driven phonon mode in a non-linear lattice results in the formation of equi-spaced spectral lines at a characteristic frequency $\Delta\omega$ set by the drive frequency and the intrinsic phonon mode frequencies. Mathematically, the DNR phenomenon can be modelled through the coupled dynamics.

$$\ddot{Q}_1 = -\omega_1^2 Q_1 - 2\zeta_1\omega_1\dot{Q}_1 + f_d\cos(\omega_d t) + \alpha_{11}Q_1^2 + \alpha_{22}Q_2^2 + \beta_{111}Q_1^3 + \beta_{122}Q_1Q_2^2 \qquad (1\text{-}1)$$

$$\ddot{Q}_2 = -\omega_2^2 Q_2 - 2\zeta_2\omega_2\dot{Q}_2 + \alpha_{12}Q_1Q_2 + \beta_{112}Q_1^2 Q_2 \qquad (1\text{-}2)$$

where $f_d$ is the displacement or drive level of drive tone $\omega_d$, $\alpha_{ij}$ & $\beta_{ijk}$ are quadratic and cubic coupling coefficients and $\zeta_{i=1,2}$ are the damping coefficients. When the frequency of external driving $\omega_d$ matches the resonant mode frequency $\omega_1$ and at high enough values of drive, the resonant mode $\omega_d$ and auto-parametrically triggered sub-harmonic mode $\omega_d/2$ are only excited with no additional spectral lines. Once the drive frequency $\omega_d$ is set beyond the dispersion band, at high enough values of drive $f_d$, the mode $Q_2$ is auto-parametrically triggered at the frequency $\omega_1/2$. This in turn results in a response for $Q_1$ at $\omega_1$ and through higher order nonlinear coupling between $Q_1$ and $Q_2$ defined in the equation 1, the near-resonant terms of $\cos(\omega_1 + p(\omega_d - \omega_1)); p \, \epsilon \, Z$ are generated (See supplementary section S1 for the analysis).

A microscopic 1-D extensional mode resonator of dimensions 1100 µm x 350 µm x 11 µm represents the experimental platform for this work (Supplementary figure S1). The spectrum analyser (SA) and Laser Doppler Vibrometry (LDV) (Supplementary figure S1) measurements prove the existence of an auto-parametrically generated sub-harmonic mode [16] (Figure 1A) upon the drive level crossing a specific threshold value. Here, the displacement profile corresponding to the sub-harmonic mode can be conceived as a pre-stressed framework for the level of coupling between drive frequency and intrinsic resonance mode (Figure 1B). That being said, the propensity for comb generation is higher at the antinodes of sub-harmonic mode. Additionally, figure 1b1 provides an evidence for the phase coherency of equidistant comb lines. In this letter, we systematically report the experimental results carried out in the extensional resonator based test bed to understand the frequency comb generation and discuss the opportunities for active tuning of comb structure.

The drive amplitude dependence on the phononic comb for an off-resonant drive frequency of 3.862 MHz is examined first (Figure 2A). Mere drive tone is present at low enough drive power levels <3 dBm. However, upon the drive power level meeting a certain threshold value of ~3.5 dBm, the comb eventually gets generated. Henceforth, with increase in drive power, the comb structure extends into higher orders. The steps in figure 2A depict the existence of inherent thresholds for each higher order spectral line. The dependence of drive frequency on the comb generation thresholds is presented in the figure 2B. When the drive frequency is operated within the dispersion band (~3.8590 MHz) (<3.8602 MHz), the comb is not formed at even extreme power levels as high as 23.99 dBm. This exemplifies the absence of the comb generation pathway in the 'dispersion band' regime. Nonetheless, outside this range > 3.8602 MHz, the drive power level threshold and drive frequency follow a direct relationship, corresponding to the Lorentzian decay of resonant peak. Figures 2a-2g show the underlying spatial manipulation of phonons from the spectral line $\omega_d$ to comb lines. Spatially, the phonons at the comb frequencies are present at the antinodes of auto-parametrically driven sub-harmonic mode with a concomitant reduction in the drive phonon population at these spatial locations.

Under high enough drive amplitude conditions (>~5 dBm), there is an evidence for Duffing nonlinear mechanism or foldover effect (Figure 3A) [12-14]. Here, the frequency of the driven mode is amplitude dependent which results in increased comb spacing corresponding to the drive amplitude (Figure 3A). To study the dependence of 'detuning' of drive tone on both DNR and Duffing influenced DNR regimes, the underlying frequency responses have to be juxtaposed. Yet, at a constant drive power level, the number of comb lines produced can be dependent on the 'detuning' level of drive. Hence, the nature of comb structure for different 'detuning' levels cannot be directly compared. Alternatively, by standardizing to 5 comb lines, the resulting frequency contours are obtained (Figures 3B and 3D). This standardization, however, results in Duffing phenomenon at higher drive frequencies >3.8622 MHz as the higher threshold for reaching 5 comb lines directly meets the Duffing criterion (Figures 3B and 3D). In DNR, the far-detuned drive frequency results in larger comb spacing (Figure 3D). In contrast, the comb spacing in Duffing influenced DNR remains the same and the overall comb collectively gets shifted with the drive frequency shift. This crossover can be explained by the fold-over of resonant peak resulting in frequency renormalization ($\widetilde{\omega}_0$) [11]. The influence of Duffing mechanism on the nominal DNR is guided by the intrinsic Duffing and DNR thresholds. An active tuning of these attributes in a microstructure [17] will enable independent control over both DNR and Duffing mechanisms and consequent tailoring of the frequency comb in both regimes. In figure 3A, there is an additional evidence for the 'filling-in' of multiple spectral lines within the comb (Cyan regions) in the Duffing influenced DNR regime. This intriguing high-generation comb phenomenon needs investigation, despite that this result falls out of the scope of this letter.

While the individual amplitude and frequency dependences are explored in both DNR and DNR-Duffing regimes, the interplay between these two parameters also reveals an additional intriguing property. As seen before, the frequency detuning threshold for frequency comb generation is dependent on the dispersion bandwidth. Within the dispersion band, the comb is not generated. Outside the band, the amplitude required for comb generation continuously increases as the drive frequency is increasingly spaced away from resonance. This clearly signifies that the quality factor of resonance peak and resonant trans-conductance are strong determinants for frequency comb generation. A microstructure with tuneable quality factor and trans-conductance [18] can therefore enable active control over the dynamics of phononic comb generation.

Figure 4A maps the 'Dispersion band', 'Phononic comb' and 'Small signal' regimes in the driving response. For studying the underlying spatial aspects in each of these regimes, the RMS surface displacement averaged over the frequency range 3.6 to 4.2 MHz (using LDV measurements) are compared. Figure 4B shows the RMS vibration patterns for different drive frequencies. In the dispersion band (at 3.858 MHz), since there is no comb formation, the resonant mode shape is not altered. Contrary to this, in the phononic comb regime (at 3.856 and 3.862 MHz), the sub-resonant mode pattern is observed following the phonon manipulation from the drive frequency to comb spectral lines (Figures 2a-2g). In the 'Small signal regime', similar to the dispersion band, the comb is not formed at normal drive levels and the resonant mode pattern is preserved. To understand the relevance of drive power levels on the spatial aspects of comb formation, the underlying RMS vibration patterns corresponding to drive frequency 3.862 MHz are compared. As seen in the figure 4C, the RMS vibration patterns gradually shift from the resonant to sub-harmonic mode pattern following phonon manipulation.

In summary, this letter reports the first ever experimental demonstration of a phononic frequency comb. Further, a phenomenological model has been specified to define the nature of direct nonlinear resonances that govern frequency comb generation. The presented concepts find general relevance to other nonlinear systems in both quantum and classical domains. Besides the fundamental advance reported here, phononic frequency combs also find applications to accurate MEMS/NEMS resonant sensors adapted for stable long duration measurements [19-20], engineering phase-coherent phonon lasers [21], phonon computing [22-23], pulse train mediated ultrasonic imaging and fundamental investigations of nonlinear phononics [24]. Further, the combination of both optical and phononic frequency combs can also enable the extension of metrology link between optical frequencies to sub-Hz frequencies.

**Acknowledgements**

Funding from the Cambridge Trusts is gratefully acknowledged.

**References**

1. Udem, Th, Holzwarth, Ronald, and Hansch, Theodor W., Optical frequency metrology. *Nature* **416** (6877), 233 (2002).

2. Del'Haye, P. et al., Optical frequency comb generation from a monolithic microresonator. *Nature* **450** (7173), 1214 (2007).

3. Kippenberg, Tobias J., Holzwarth, Ronald, and Diddams, S. A., Microresonator-based optical frequency combs. *Science* **332** (6029), 555 (2011).

4. Ye, Jun, Schnatz, Harald, and Hollberg, Leo W., Optical frequency combs: from frequency metrology to optical phase control. *Selected Topics in Quantum Electronics, IEEE Journal of* **9** (4), 1041 (2003).

5. Holzwarth, R. et al., Optical frequency synthesizer for precision spectroscopy. *Physical review letters* **85** (11), 2264 (2000).

6. Jones, R. Jason and Diels, Jean-Claude, Stabilization of femtosecond lasers for optical frequency metrology and direct optical to radio frequency synthesis. *Physical review letters* **86** (15), 3288 (2001).

7. Jones, David J. et al., Carrier-envelope phase control of femtosecond mode-locked lasers and direct optical frequency synthesis. *Science* **288** (5466), 635 (2000).

8. Thorpe, Michael J. et al., Broadband cavity ringdown spectroscopy for sensitive and rapid molecular detection. *Science* **311** (5767), 1595 (2006).

9. Abbott, B. P. et al., Observation of gravitational waves from a binary black hole merger. *Physical review letters* **116** (6), 061102 (2016).

10. Kippenberg, T. J., Spillane, S. M., and Vahala, K. J., Kerr-nonlinearity optical parametric oscillation in an ultrahigh-Q toroid microcavity. *Physical review letters* **93** (8), 083904 (2004).

11. Cao, L. S. et al., Phononic Frequency Combs through Nonlinear Resonances. *Physical review letters* **112** (7), 075505 (2014).

12. Kovacic, Ivana and Brennan, Michael J., *The Duffing equation: nonlinear oscillators and their behaviour*. (John Wiley & Sons, 2011).

13. Aldridge, J. S. and Cleland, A. N., Noise-enabled precision measurements of a Duffing nanomechanical resonator. *Physical review letters* **94** (15), 156403 (2005).

14. Eichler, A. et al., Nonlinear damping in mechanical resonators made from carbon nanotubes and graphene. *Nature nanotechnology* **6** (6), 339 (2011).

15. Fermi, Enrico, Pasta, J., and Ulam, S., Studies of nonlinear problems. *Los Alamos Report LA-1940* **978** (1955).

16. Chechin, G. M., Novikova, N. V., and Abramenko, A. A., Bushes of vibrational modes for Fermi-Pasta-Ulam chains. *Physica D: Nonlinear Phenomena* **166** (3), 208 (2002).

17. Kozinsky, I., Postma, H. W. Ch, Bargatin, I., and Roukes, M. L., Tuning nonlinearity, dynamic range, and frequency of nanomechanical resonators. *Applied Physics Letters* **88** (25), 253101 (2006).

18. Verbridge, Scott S., Shapiro, Daniel Finkelstein, Craighead, Harold G., and Parpia, Jeevak M., Macroscopic tuning of nanomechanics: substrate bending for reversible control of frequency and quality factor of nanostring resonators. *Nano Letters* **7** (6), 1728 (2007).

19. Middlemiss, R. P. et al., Measurement of the Earth tides with a MEMS gravimeter. *Nature* **531** (7596), 614 (2016).

20. Cleland, Andrew N. and Roukes, Michael L., A nanometre-scale mechanical electrometer. *Nature* **392** (6672), 160 (1998).

21. Vahala, Kerry et al., A phonon laser. *Nature Physics* **5** (9), 682 (2009).

22. Wang, Lei and Li, Baowen, Thermal logic gates: computation with phonons. *Physical review letters* **99** (17), 177208 (2007).

23. Stannigel, K. et al., Optomechanical quantum information processing with photons and phonons. *Physical review letters* **109** (1), 013603 (2012).

24. Forst, Michael et al., Nonlinear phononics as an ultrafast route to lattice control. *Nature Physics* **7** (11), 854 (2011).

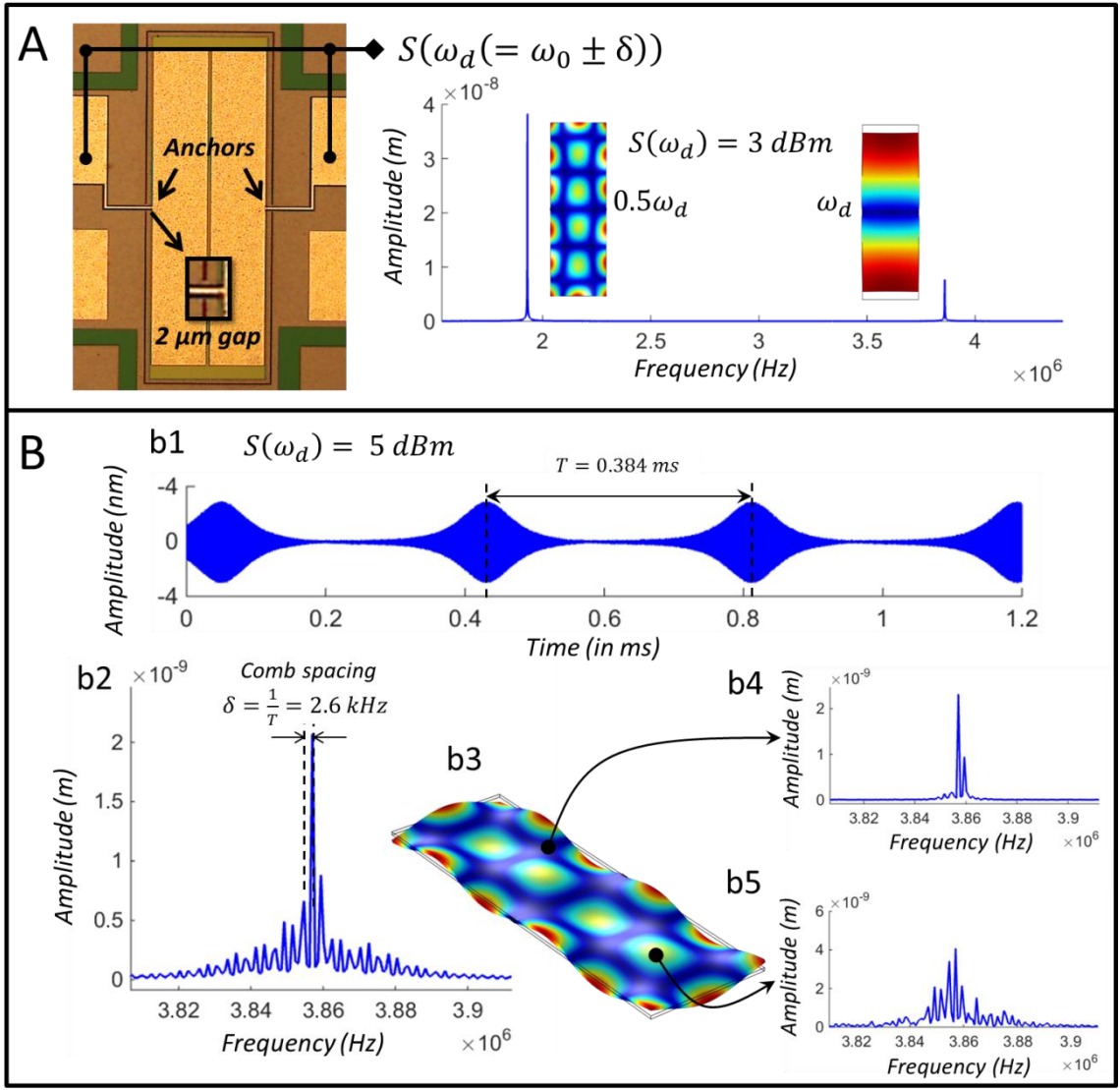

Figure 1: **Observation of phononic frequency comb.** A: Left: Signal $S(\omega_d)$ is applied on a free-free beam microstructure; Right: An intrinsic parametric excitation of sub-harmonic mode (out-of-plane) at the drive power level 3 dBm of in-plane extensional mode; B: b1: The pulse train corresponding to the phase coherent frequency comb at the drive power level 5 dBm; b2: Surface average displacement spectrum demonstrating comb formation; b3: The sub-harmonic mode shape; b4: The displacement at the antinode of sub-harmonic mode indicating the absence of the frequency comb response; b5: The displacement at the node of sub-harmonic mode indicating the presence of the frequency comb response;

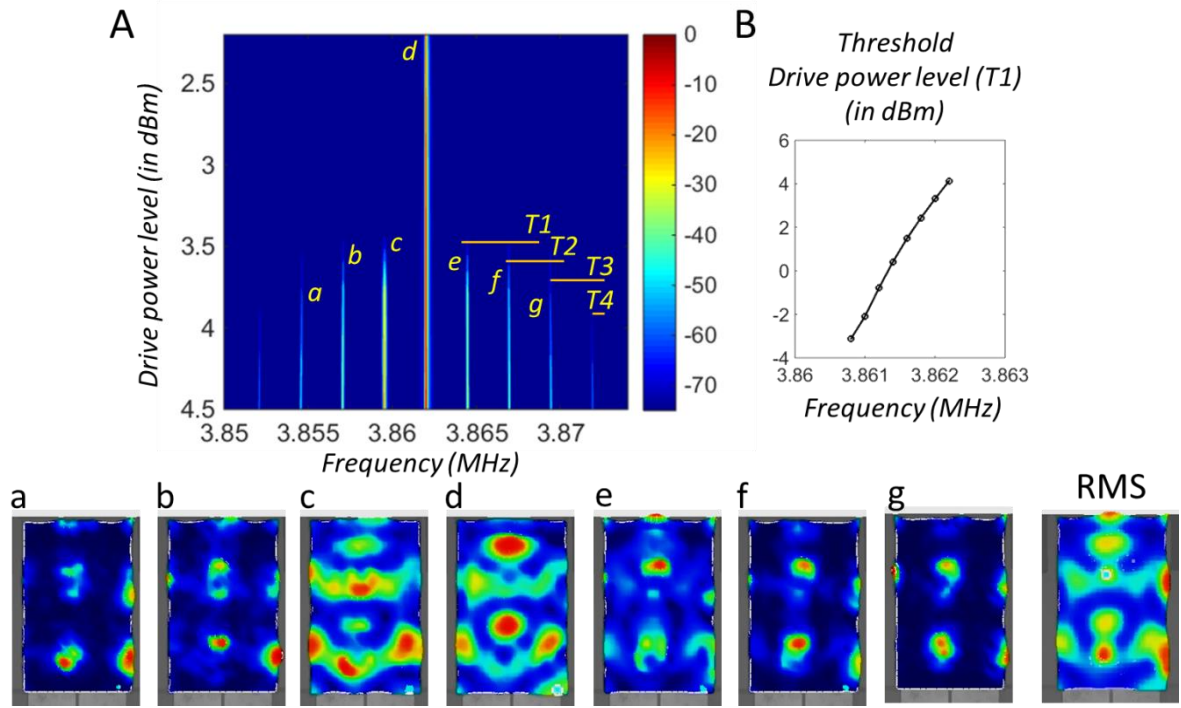

Figure 2: **High-order phononic frequency combs.** A: The drive power level-frequency contour indicating the thresholds $T_1$, $T_2$, $T_3$, $T_4$ for high-order frequency comb generation; a-g: The out-of-plane displacement profiles at different frequencies on both the sides of dispersion curve; RMS: The RMS displacement profiles in the frequency range 3.6-4.2 MHz at the drive frequency 3.862 MHz and drive power level = 4.5 dBm; B: The threshold $T_1$ as the drive frequency is detuned from resonance.

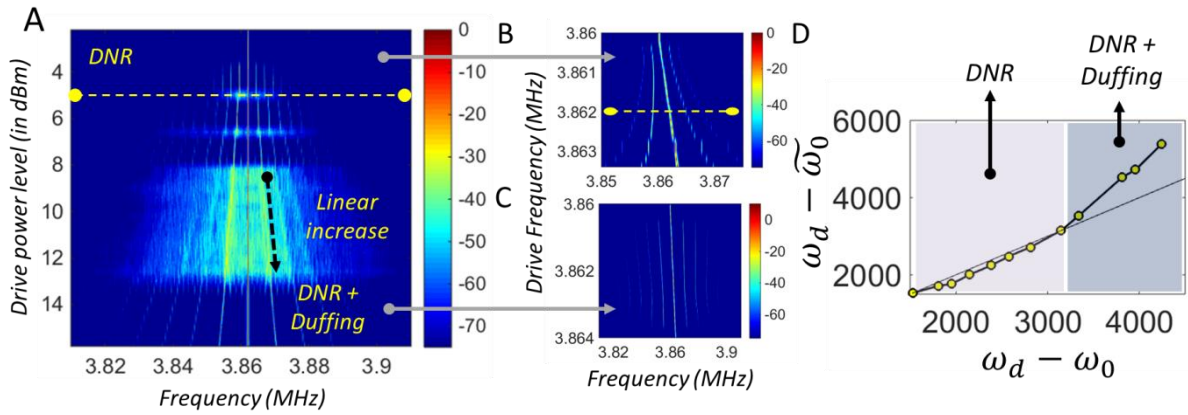

Figure 3: **Interplay between DNR and Duffing mechanisms.** A: The drive power level-frequency contour (drive frequency = 3.862 MHz) indicating the DNR and DNR-Duffing regimes; B and C: The drive frequency-sense frequency contour (5 comb lines) in the DNR and DNR-Duffing regimes respectively; D: The comb spacing with frequency renormalization ($\omega_d - \widetilde{\omega_0}$) vs. the comb spacing without frequency renormalization ($\omega_d - \omega_0$) indicating the DNR and Duffing regimes.

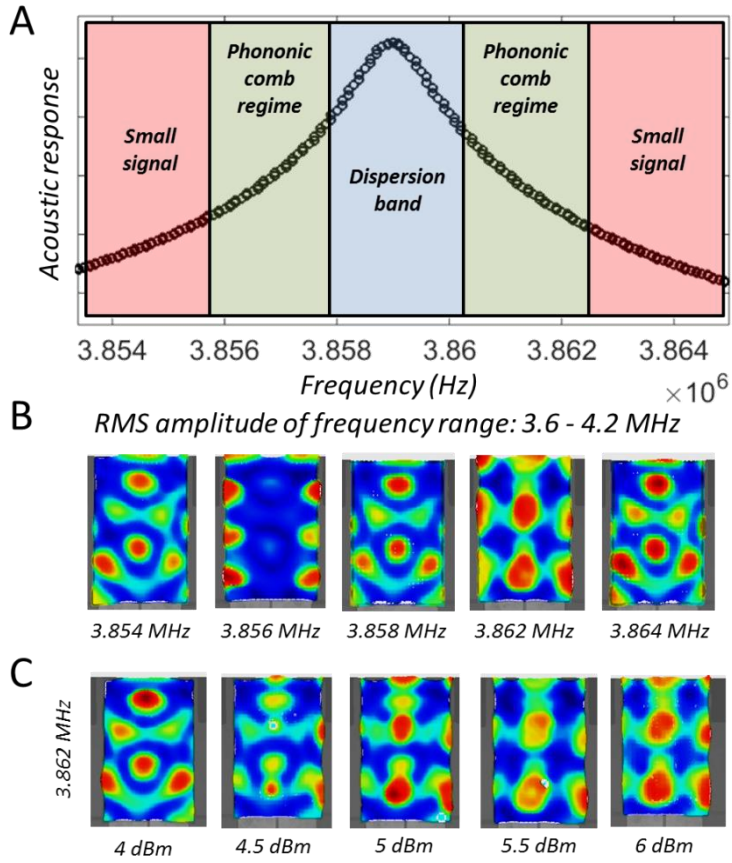

Figure 4: **Regime of phononic comb formation.** A: Phononic comb regime falling just outside the dispersion band of the mechanical mode; B and C: The RMS displacement profiles in the frequency range 3.6-4.2 MHz at different drive frequencies (drive power level = 6 dBm) and drive amplitudes ($\omega_d$=3.862 MHz) respectively.

**Supplementary Information**

**Phononic frequency comb via intrinsic three-wave mixing**

Authors: Adarsh Ganesan[1], Cuong Do[1], Ashwin Seshia[1]

1. Nanoscience Centre, University of Cambridge, Cambridge, UK

## 1. Analytical derivation of DNR frequency comb

For a truncated phase space spanned by two phonon modes $Q_1$ & $Q_2$, the dynamics of FPU chain can be written as,

$$\ddot{Q}_1 = -\omega_1^2 Q_1 - 2\zeta_1\omega_1\dot{Q}_1 + f_d \cos(\omega_d t) + \alpha_{11}Q_1^2 + \alpha_{22}Q_2^2 + \beta_{111}Q_1^3 + \beta_{122}Q_1Q_2^2 \quad \text{(S1-1)}$$

$$\ddot{Q}_2 = -\omega_2^2 Q_2 - 2\zeta_2\omega_2\dot{Q}_2 + \alpha_{12}Q_1Q_2 + \beta_{112}Q_1^2 Q_2 \quad \text{(S1-2)}$$

where $f_d$ is the displacement or drive level of drive tone $\omega_d$ and $\alpha_{ij}$ & $\beta_{ijk}$ are quadratic and cubic coupling coefficients of FPU chain. Note: The damping terms are dropped off for clarity.

The 0$^{\text{th}}$ order solution of $Q_1$ will be just a linear oscillator with the tone $cos(\omega_d t)$. The $Q_1^{(0)}$ solution is now coupled into the dynamics of $Q_2$ to obtain $Q_2^{(0)}$. This case presents a Mathieu framework. Outside the dispersion band, the parametric loading offered by $Q_1$ on $Q_2$ triggers the mode at frequency $\omega_2 \approx \omega_d/2$. The inter-modulation between drive tone and eigenmode is established through quadratic and cubic nonlinear terms in (S1-1) and the tones $cos(\omega_1 t)$ and $cos((\omega_d - \omega_1)t)$ are produced in the truncated frequency spectrum of $Q_1^{(1)}$. By solving equation (S1-1) though sequential iteration the coupling between tones at $(\omega_d - \omega_1)$ and $\omega_1$ is established through the quadratic and cubic nonlinear terms in (S1) and the near-resonant terms of $cos(\omega_1 + p(\omega_d - \omega_1)t); p \in Z$ are generated. In addition, the tones at $cos\left(\frac{\omega_1}{2} + p(\omega_d - \omega_1)t\right)$ and $cos(p(\omega_d - \omega_1)t); p \in Z$ are generated.

Supplementary figure S1

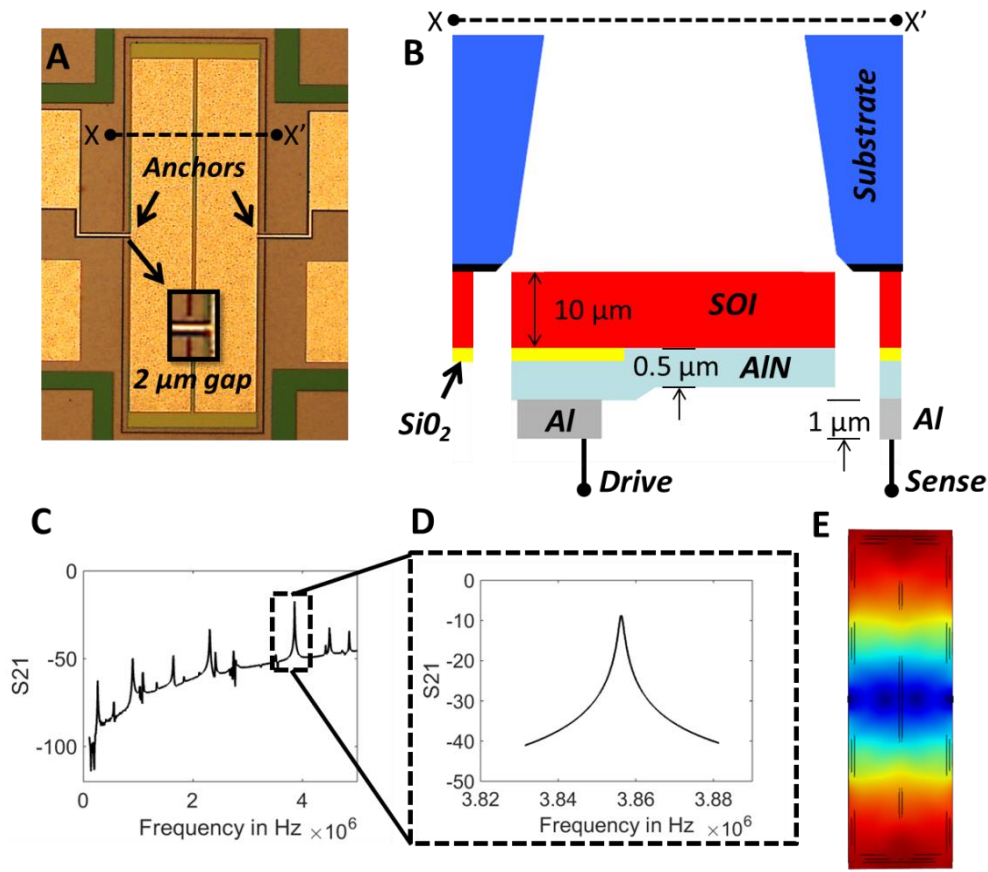

**Figure S1: Operation of piezoelectrically driven micromechanical resonator**: **A**: Free-free beam topology with 2 μm air gap for in-plane mode excitation; **B**: 1 μm thick Al electrodes patterned on 0.5 μm thick AlN piezoelectric film which is in-turn patterned on SOI substrate; the 10 μm thick SOI layer is then released through back-side etch to realize mechanical functionality; **C**: The scattering parameter S21 denoting forward transmission gain across a broad range of frequencies from 0-4.5 MHz; **D**: across a smaller range of frequencies 3.83-3.88 MHz; **E**: Resonant mode shape of ~3.86 MHz vibrations from eigen-frequency analysis in COMSOL and Doppler shift vibrometry.

Supplementary figure S2

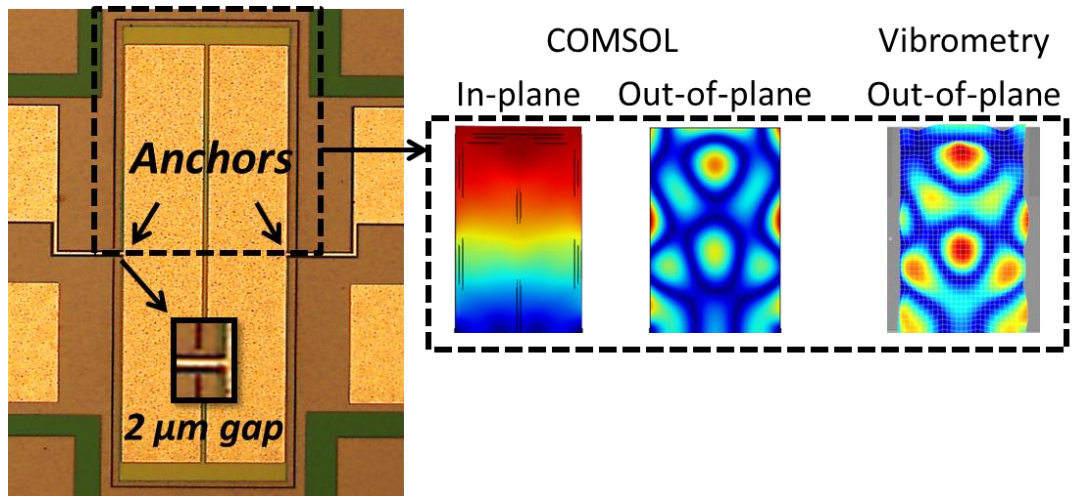

**Figure S2:** Mode shape prediction using eigenfrequency analysis in COMSOL and Doppler shift vibrometry measurement